\documentclass[]{BioPhysMath}
\usepackage[OT4,T1]{fontenc}
\usepackage[utf8]{inputenc}
\usepackage{polski}
\usepackage{graphicx}

\usepackage{amsmath}
\usepackage{amssymb}

\usepackage[english,polish]{babel} 
\usepackage{lastpage} 

\usepackage{color} 
\usepackage{cite} 
\usepackage{tabularx}
\RequirePackage[numbers]{natbib}

\usepackage[colorlinks=true]{hyperref}
  \hypersetup{
    pdftitle={Template BioPhysMath}, 
    pdfauthor=Nowy Autor, 
    colorlinks,
    urlcolor=blue,
    filecolor=magenta,
    citecolor=green, 
    linkbordercolor={1 1 1}, 
    citebordercolor={1 1 1},  
    urlbordercolor={ 1 1 1}  
  } 

\newcommand{\orcid}[1]{\href{https://orcid.org/#1}{\includegraphics[scale=.05]{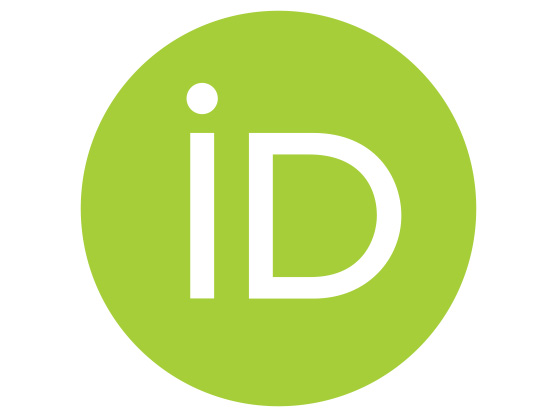}}}

\pdfoutput=1
\usepackage{graphicx}
\usepackage{wrapfig}
\graphicspath{{./Figures/},{./Pictures/}}
\usepackage{multicol}
\firstpage{i}    
\LogoG{\vspace*{1cm}\includegraphics[width=0.1\textwidth]{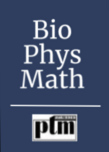}}
\volume{2}
\fasc{1}
\years{2024}
\LogoGMS{\includegraphics[width=0.18\textwidth]{bpm.png}}
\wwwtrue

\pages{24}
\received[7th May 2024]{30th January 2024}
\lastrevision{}
\logo{}{}{}
\def\doinum{10.14708/bpm.v43i1.xxx}

\title[Forest insect outbreaks ]{How forest insect outbreaks depend on forest size and tree distribution: an individual-based model results
}

\tpauthortrue 
\author[J. Uchmański]{Janusz Uchmański\orcid{0000-0001-8087-8371}}
\thanks{J. Uchmański}
\affiliation{Cardinal Stefan Wyszyński University in Warsaw}
\address{Wóycickiego 1/3,
01-938 Warsaw,
Poland}
\email{j.uchmanski@uksw.edu.pl}


\begin{document}
\vspace{-10ex}
\renewcommand{\thefootnote}{}
\footnote{\href{http://creativecommons.org/licenses/by/3.0/}{Licensed under a Creative Commons Attribution License (CC-BY)}}
\setcounter{page}{1} 
\selectlanguage{english}\Polskifalse

\begin{abstract}
In this work, an individual-based model of forest insect outbreaks is presented. The results obtained show that the outbreak is an emerging feature of the system. It is a common product of the characteristics of insects, the environment in which the insects live, and the way insects behave in it. The outbreak dynamics is an effect of scale. In a sufficiently large forest regardless of the density of trees and their spatial distribution, provided that the range of insect dispersion is large enough, it develops in the form of an outbreak. In very small forests, the dynamics becomes more chaotic. It loses the outbreak character and, especially in the forest with random tree distribution, there is a possibility that the insect population goes extinct. The local dynamics of the number of insects on one tree in a forest, where the dynamics of all insects has the character of outbreak, is characterized by a rapid increase in number and then a rapid decrease until the extinction of the local population. It is the result of the influx of immigrants from neighboring trees. The type of tree distribution in the forest becomes visible when the density of trees becomes low and/or the range of insect dispersion is small. When trees are uniformly distributed and the range of insect dispersion is small, the system persists as a set of more or less isolated local populations. In the forest with randomly distributed trees, the insect population becomes more susceptible to extinction when the tree density and/or range of insect dispersion are small.
\end{abstract}

\section{Introduction}
\vskip 0.3cm

Forest insect outbreaks are an intensively investigated population phenomenon \cite{morris, barbosa, berryman, berryman2, isaev2}. Classical models describing forest insect outbreaks from a mathematical point of view are systems of differential equations. Their analysis comes down to searching for conditions in which the solutions of these equations are characterized by oscillations \cite{ludwig, nedorezov, hassel, isaev}. The state variable in these equations is the density of the insect population. When interpreting the results of these models and comparing them with actual data on insect numbers or densities from forest measurements, it is difficult to decide what the model results should be compared to. Classic models of these phenomena most often do not take into account the spatial structure of the modeled object. Presenting this problem in a slightly caricatured way, it can be said that classical models seem to describe the resources used by insects as one huge tree whose biomass is equal to the sum of the biomass of all trees in the forest. Then, the state variable used in the model will, of course, be the total density of all insects in the forest, i.e.~the total number of insects in the forest divided by the area covered by the forest. The situation becomes absurd because such a~value cannot be measured in a real forest and it is not known whether insect density measurements made on a smaller spatial scale accurately reflect what we want.

Some models take into account the spatial structure of the modeled system. Partial differential equations with diffusion terms are then used (recently, for example, \cite{shu}). This creates another type of difficulty in interpreting the results. The use of partial differential equations means that space is treated in a continuous manner. So, what should we compare the point values of insect density and the amount of resources with? The actual system has a clearly discontinuous spatial structure.

Individual-based models create different possibilities to interpret their results. In an early published individual-based model that describes the dynamics of forest insects, the conditions that favor the appearance of outbreaks were analyzed \cite{uchmanski3}. This model tracked the fate of each insect and the biomass dynamics of each tree in the forest. However, all results were obtained for very specific spatial conditions in which the system operates. It was a forest, or rather a plantation, in the shape of a square with evenly planted trees. 20 trees were planted along each side of this square. So in total there were 400 trees in this forest. The question arises how the dynamics of the system depends on the size of the area on which the forest grows and what impact other types of distribution of trees in the space of forest have on the dynamics of the system.

Ludwig \cite{ludwig2} considered a similar problem of what is the critical size of a forest patch that can support an outbreak. However, they applied the classical approach to solve this problem. These were nonlinear differential equations with diffusion terms. Therefore, it is difficult to compare their results with the results of the individual-based model presented in this work.

The model described in this and previous \cite{uchmanski3} papers does not refer to any real population, but during construction of it I had in mind \textit{Panolis flammea} outbreaks in pine forests in western Poland \cite{kolk}.

\section{Basic model – forest size 20x20, uniform tree distribution
} 
\subsection{The model}
\vskip 0.3cm
The forest stand consists of 400 trees uniformly distributed in a two-dimensional square space. Each side of this square contains 20 trees. To avoid the edge effect, the opposite sides of the square are in contact with each other. The area covered by the forest was divided into evenly arranged patches where the trees could potentially grow. The actual size of such a spatial patch may correspond to the size of the tree's root system or be the projection of the tree crown onto the ground surface. This allows for comparison of events in forests with different tree densities and different types of spatial tree distribution. The size of a patch that accommodates one tree is also a~linear measurement unit in this model. 

The larvae of an insect species whose life span is one year consume tree resources. The rate of assimilation of the resource $A$ by the larva and the living costs of this larva, measured by the respiration rate $R$, are power functions of its body weight \cite{duncan}:

\begin{equation}\label{1}
A=a_1w^{b_{1}},
\end{equation}  
\begin{equation}\label{2}
R=a_2w^{b_{2}},
\end{equation}
where  $w$ is the body weight of a larva, $a_1$, $a_2$, $b_1$ and $b_2$ are parameters. Equations \eqref{1} and \eqref{2} give the following equation of the larval growth:
\begin{equation}\label{3}
\frac{dw}{dt} = a_1 w^{b_{1}} - a_2 w^{b_{2}}.
\end{equation}

  The rate of assimilation by a larva also depends on the amount of available resources. The rate of assimilation $A$ by a single larva, isolated from interactions with other larvae of the same species, as a function of the available resources $V$ and weight $w$, can be described by the Ivlev’s equation \cite{ivlev}:
  \begin{equation}\label{4}
  A=a_{1,\max}(1-\text{e}^{sV})w^{b_{1}},
  \end{equation}
where $a_{1,\max}$ and $s$ are constant parameters. But larvae living together on a~tree compete for resources \cite{balten}. This leads to unequal partitioning of resources among competitors \cite{lomnicki}. The rate of assimilation of a larva in a local population of competing larvae on a given tree is described by eq. \eqref{4} with additionally introduced dependence on the current body weight of the larva \cite{uchmanski2, uchmanski5} according to the following rule. In each time step during the simulation of larva growth, an individual with the lowest $w_{\min}$ and the highest $w_{\max}$ body weight was determined in each local population. The value of the parameter $a_1$ for the lightest larva is described by
\begin{equation}\label{5}
a_{\min}=a_{1,\max}(1-\text{e}^{s_{\min}V})
\end{equation}
  and that of the heaviest larva by
\begin{equation}\label{6}
a_{\max}=a_{1,\max}(1-\text{e}^{s_{\max}V}),
\end{equation}
  where $V$ is the actual amount of tree resources. The parameter $a_1$ of larvae with intermediate weights is calculated using linear interpolation between the values $a_{\min}$ for $w_{\min}$ and $a_{\max }$ for $w_{\max}$. The assimilation rate calculated this way separately for each competing larva in the population was used in eq.~\eqref{3} to calculate its weight increase in each simulation step.

The values of constant parameters $s_{\min}$ and $s_{\max}$ of eqs.~\eqref{5} and \eqref{6} fulfil the following inequality:
\begin{equation}\label{7}
s_{\min}\leq s_{\max}.
\end{equation}
  For $s_{\min}  = s_{\max }$ resource assimilation is the same for all competing larvae. Its values increase with increasing $V$. For $s_{\min} < s_{\max}$ larvae differ in the rate of assimilation. The greater the difference between $s_{\min}$ and $s_{\max}$, the greater is the individual variability. However, this variability disappears for $V \to \infty $.  This algorithm describing resource partitioning between competing individuals has been supported by many experiments and observations of the growth of competing individuals in even-aged populations and their weight distributions \cite{uchmanski}.

A larva growing according to the equation
\begin{equation}\label{8}
\frac {dw_{\max}} {dt} = a_1w_{\max}^{b_{1}} - a_2w_{\max}^{b_{2}}
\end{equation}
  has the highest weight $w_{\max}$ in successive time steps of the growth simulation and at the end of growth. The final weight $w_{\max}^{\text{end}}$ reached asymptotically when the assimilation equals respiration for the growth described by eq.~\eqref{8} is equal to:
\begin{equation}\label{9}
w_{\max}^{\text{end}}= \left( \frac {a_{1,\max}} {a_2} \right)^{\frac {1} {b_2 - b_1}}.
\end{equation}
  The final weight mentioned above was well approximated if the growth of the larvae according to eq.~\eqref{8} was numerically solved in 80 time steps. The growth equations for other larvae were also solved numerically in 80 time steps.

The larvae started feeding on trees in spring. Their growth was terminated in autumn. At the initial year $N_{0}$ larvae start feeding on a tree located in the middle of the space. Their initial body weights are in the range $[{w}_{0,\min}, \textit{w}_{0,\max }]$ and are derived from a normal distribution with a mean value $w_{0, \text{mean}}$ and a variance ${w}_{0,\text{variance}}$. At each time step, the highest and lowest body weights of the larvae were found on each tree, so that it was possible to calculate the consumption of each larva with respect to their variability in the local population, which depends on the actual level of resources available in the tree.

At each time step, the body weight of each larva was compared with the highest possible weight at this time step, obtained from eq.~\eqref{8}. If it was less than $mort\_{juvenil}\, w_{\max}$ (where $mort\_{juvenil} $ is a parameter: $0 \leq mort\_{juvenil}  < 1$), this larva died and was removed from the population.

The larval growth was stopped after 80 time steps. The larvae became adult individuals with weights equal to those at the end of growth. It was assumed that adult individuals do not feed. The variability in adult body weight was related to the food conditions under which the larvae were growing. An adult individual with probability $d_{0}$ ($0 \leq {d}_{0} < 1$) can remain on the tree where it foraged as a larva and lay eggs here, or with probability $1 - {d}_0$ move to one of the neighboring trees to lay eggs there. The trees were uniformly distributed at the nodes of the quadratic network. Dispersion could occur along one of the four directions with equal probability: vertical up or down, and horizontal to the left or to the right. The distance $d$ of dispersion, measured along these directions in units of the smallest distance between two neighboring trees, was taken from the following distribution:
\begin{equation}\label{10}
d=1+\mathrm{round}(\mathrm{abs}(RandG)),
\end{equation}
  where $\mathrm{round}$ is a function rounding a real number to the nearest integer, $\text{abs}$ is the absolute value of the number, and $RandG$ is a random number drawn from a normal distribution with mean of 0 and standard deviation $disp_{sd}$. 

The probability of death of an adult individual was described by parameter ${mort}\_{adult}$ ($0\leq {mort}\_{adult} < 1$). Adult individuals which survived will reproduce. 

The model describes the population dynamics of a parthenogenetic species. The number of progeny $z$ produced by an adult insect was proportional to the difference between the final weight of the larva $w_{\mathrm{end}}$ and a threshold weight $w_{\mathrm{fak}}\,w_{\max}^{\mathrm{end}}$:

\begin{equation} \label{eq11}
z=\left\{
\begin{tabular}{ccc}
$\mathrm{round}(c(w_{\mathrm{end}}-w_{\mathrm{fak}}w_{\max}^{\mathrm{end}}))$ & \textrm{for} &   $w_{\mathrm{end}}>w_{\mathrm{fak}}\, w_{\max}^{\mathrm{end}}$,\\
0 & \textrm{for}  &  $w_{\mathrm{end}} \leq w_{\mathrm{fak}}\, w_{max}^{\mathrm{end}}$,
\end{tabular}
\right.
\end{equation}
where $c$ is a parameter that describes the intensity of progeny production, and $w_{\mathrm{fak}}$ ($0 < w_{\mathrm{fak}} < 1$) shows the proportion of the threshold weight which an individual needs for progeny production to the maximum final weight $w_{\max}^{\mathrm{end}}$ given by eq.~\eqref{9}. Adult insects with weights lower than or equal to the threshold weight die without reproduction. The function $\mathrm{round}$ in Eq.~\eqref{eq11} rounds a real number to the nearest integer. The initial weights of the progeny produced by each individual were drawn from a normal distribution with mean $w_{0, \mathrm{mean}}$ and variance $w_{0,\mathrm{variance}}$, but their values had to be in the range $[{w}_{0,\min }, {w}_{0,\max }]$. Adult insects died after reproduction. 

All calculations of growth and reproduction, and also all decisions concerning death, survival, or dispersal were calculated for each individual separately in the current generation.

The model assumes that insects live on fully developed trees, not growing any more, and equations of the rate of change in tree resources describe only the regeneration of these resources. At the initial time instant, each tree provided $V_{0}$ resources for insects. They were decreased at a rate equal to the sum of the rates of assimilation by all larvae feeding on a particular tree, and renewed at a rate logistically dependent on the actual amount of resources:
\begin{equation} \label{12}
\frac{dV}{dt}=-\sum_{i=1}^{N} A_i-r \left( 1-\frac{V}{V_0}\right),
\end{equation}
 where the summation is over all larvae feeding on a particular tree, and $r$ determines the rate of resource regeneration. Eq.~\eqref{12} was solved jointly with equations that describe the increase in weight of larvae in a tree and in the same number of time steps. 

If the amount of tree resources becomes lower than ${regen}\_{threshold} \, V_{0}$, where $0 < {regen}\_{threshold} < 1$, the tree is not capable of sustaining a local population of insects. Insects feeding on it died, and the tree needed $f$ years to regenerate its resources to the level ${regen}\_{threshold}\, V_{0}$. The return to this level occurred at the beginning of the year. If the tree was not colonized by insects, then the tree resources increased to the level $V_{0}$ until the end of this season. An adult individual who colonized a tree that did not regenerate its resources was not capable of reproduction.

Calculation of the number of progeny of each individual in the current year, their location on a particular tree in the forest, and the summation of the progeny of adult insects who decided to place progeny on particular trees in the forest ended the simulation for a given year. Simulations for the following year were initiated by finding larvae with the lowest and highest body weights for each colonized tree. It makes it possible to start the simulation of larval growth on all populated trees in the following year. The initial amount of resources available to the larvae resuming feeding on a tree next spring was equal to the amount of resources available at the end of the simulation in the preceding autumn. 

The simulation output will be shown as the total number of larvae on all trees in the forest, in the spring of each year, that is, at the time when they begin feeding. 
\vspace{-1ex}
\begin{table}[!h]
\caption{Standard values of the model parameters applied in simulations.}
 \label{table:1}
\centering
\begin{tabularx}{1\textwidth} { 
  | >{\centering\arraybackslash}X 
  | >{\centering\arraybackslash}X 
  | >{\centering\arraybackslash}X |
  }
 \hline
 Parameter & Value & Description \\
 \hline
 \ \ ${a}_{1,\max}$  & 0.11  & Maximal assimilation parameter   \\
 \hline
 \ \ ${a}_2$  &  0.03 & Respiration parameter \\
 \hline
 \ \ ${b}_1$ & 0.7 & Assimilation parameter  \\
 \hline
 \ \ ${b}_2$ & 0.9 & Respiration parameter \\
 \hline
 \ \ ${w}_{0,\min}$ & 14 & Minimal initial weight \\
 \hline 
 \ \ ${w}_{0,\max}$ & 26 & Maximal initial weight \\
 \hline
 \ \ ${w}_{0,\text{mean}}$ & 20 & Average initial weight \\
 \hline
 \ \ ${w}_{0, \mathrm{variance}}$ & 5  & Variance of initaial weight distribution \\
 \hline
 \ \ ${s}_{\min}$ & 0.00000051   & Ivlev's parameter, lightest individual \\
 \hline
 \ \ ${s}_{\max}$  & 0.00000120   & Ivlev's parameter, heaviest individual  \\
 \hline
 \ \ $c$ & 0.02 & Reproduction parameter  \\
 \hline
 \ \ ${w}_{\mathrm{fak}}$ & 0.65 & Threshold for reproduction \\
 \hline
 \ \ ${N}_0$  & 100 & Initial population number \\
 \hline
 \ \ ${V}_0$ & 8000000 & Initial resource of a tree \\
 \hline
 \ \ $r$  & 0.01 & Resource regeneration parameter  \\
 \hline
 \ \ ${regen}\_{threshold}$ & 0.55 & Threshold for resource regeneration  \\
 \hline
 \ \ $f$ & 6 & Number of years for resource regeneration \\
 \hline
 \ \ ${mort}\_{juvenil}$  &  0.30  & Juvenil mortality \\
 \hline
 \ \ ${mort}\_{adult}$ & 0.01 & Adult mortality \\
 \hline
 \ \ ${d}_0$ & 0.50 & Proportion of non-dispersing individuals \\
 \hline
\ \ ${disp}\_{sd}$ & 2.50 & Standard deviation of dispersal distance distribution \\
 \hline
\end{tabularx} 
\end{table}

\begin{figure}[!ht]
    \centering
    \includegraphics[width=\linewidth]{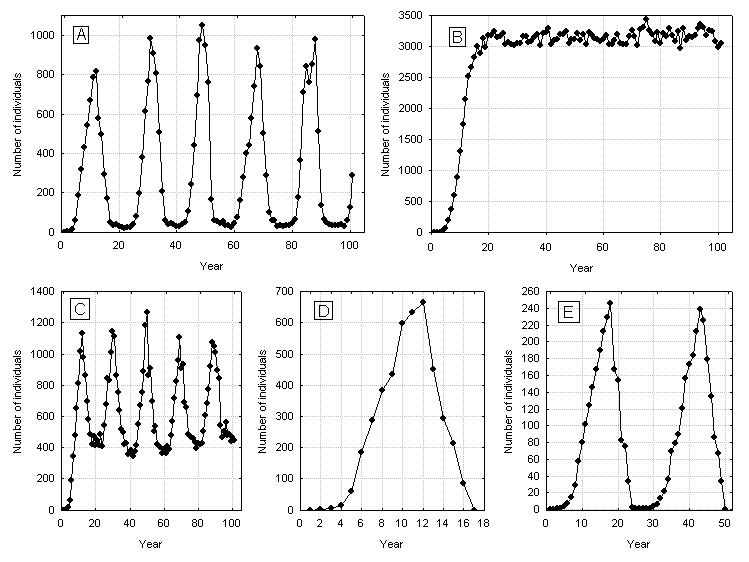}
    \caption{Examples of the typical dynamics of the total number of insects in the forest with the size $20\times 20$ (here illustrate average numbers of insects per individual tree in the forest). Parameter ${disp}\_{sd} = 0.5$. The results, except for E where $c = 0.01$, differ only in the parameters ${regen}\_{threshold}$ and $f$: A -- cyclic outbreak, ${regen}\_{threshold} = 0.57$,  $f  = 12$; B -- permanent outbreak, ${regen}\_{threshold}  = 0.1$,  $f  = 10$; C -- transitional dynamics, ${regen}\_{threshold}  = 0.45$, $f  = 16$, D -- pulse outbreak, $regen\_{threshold}  =  0.7$, $f  =  4$, E -- transitional dynamics,  $regen\_{threshold}  = 0.9$,  $f  =  6$. The other parameters have standard values (Table \ref{table:1}) (from \cite{uchmanski3}). } 
   \label{fig:1}
\end{figure}

\subsection{Results}
\vskip 0.1cm
Several characteristic types of insect population dynamics were observed (Fig.~1, \cite{uchmanski3}). Outbreak dynamics -- population size is characterized by regularly occurring high numbers, between which the numbers dropped to very low values. This is a fairly persistent type of dynamics that has not led to population extinction for hundreds or even thousands of years. Permanent outbreak dynamics -- the number increases to large values and permanently remains at this level, showing only minor fluctuations. Pulse outbreak dynamics - the population size quickly increases to a maximum, and then falls sharply and the population goes extinct after a dozen or so seasons. In addition, we will observe two types of transitional dynamics. Dynamics of a transitional type between outbreak and permanent outbreak -- the number increases to a certain level, but then quite significant, more or less regular oscillations in numbers appear, which, however, do not lead to the extinction of the population. Transitional dynamics between pulse outbreak and outbreak -- the population is characterized by several (usually two) distinct population maximums, and then at the next minimum number, the population usually goes extinct out after several dozen years.

In obtaining outbreak dynamics, both the characteristics of tree resource dynamics and the characteristics of insects are very important \cite{uchmanski3}. If a tree too easily goes into a state that does not allow it to support the local population of insects, we get a pulse outbreak type of dynamics -- extinction of insects in the entire forest, but the forest is also destroyed by insects. On the other hand, if trees are too resistant to insect exploitation, we will get permanent outbreak dynamics. Only intermediate values of tree resistance guarantee the appearance of outbreak dynamics. 
However, for this to happen, the number of years needed for the tree to regenerate its resources cannot be too small. The period of occurrence of population maximums in the case of outbreak dynamics increases with increasing number of years needed for the tree to regenerate resources. Too much offspring production is also not conducive to the emergence of outbreak dynamics. Pulse outbreak dynamics will then most likely dominate. In turn, too low offspring production leads to permanent outbreak dynamics. We will get the same results when there is too little or too much individual variability in the amount of resources obtained by individuals as a result of competition with other individuals in the local population. Intermediate values of these two parameters lead to outbreak dynamics. Mortality has another impact on the dynamics of the system. Increasing larval mortality, which in this model means eliminating poorly growing larvae, eliminates all other types of dynamics and leaves only pulse outbreak dynamics, while increasing adult mortality leads to the dominance of permanent outbreak dynamics.

\section{Forests of different sizes and different tree distributions }
\vskip 0.3cm
The simulations, the results of which are presented below, were carried out for such values of the model parameters that, for a forest with a size of $20\times 20$ trees uniformly distributed over the surface, give the population dynamics of the outbreak type. The values of these parameters are shown in Table \ref{table:1}.

\begin{figure}[!h]
    \centering
    \includegraphics[width=\linewidth]{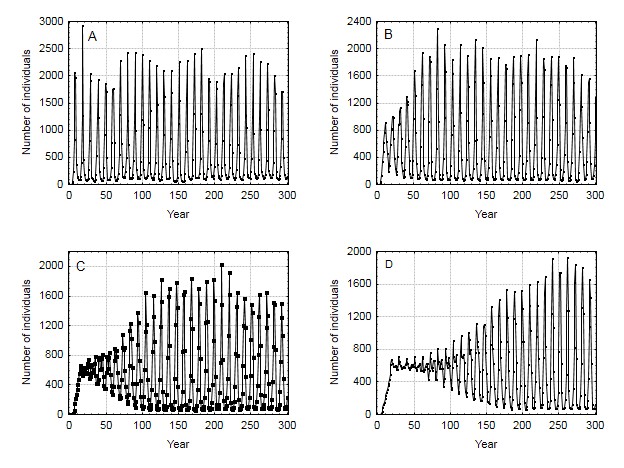}
     \caption{The development of outbreak dynamics over the initial 300 years of the simulation for forests with uniform distribution of trees in the space and different sizes. The figure shows the total number of insects in the forest divided by the number of trees in the forest. A -- forests size $20\times 20$. B -- forests size $50\times 50$. C -- forests size $75\times 75$. D -- forests size $100\times 100$. The parameters of the model have standard values (Table \ref{table:1}).}
     \label{fig:2}
\end{figure}

\subsection{Uniform distribution of trees, different forest sizes }
\begin{figure}[!h]
    \centering
    \includegraphics[width=\linewidth]{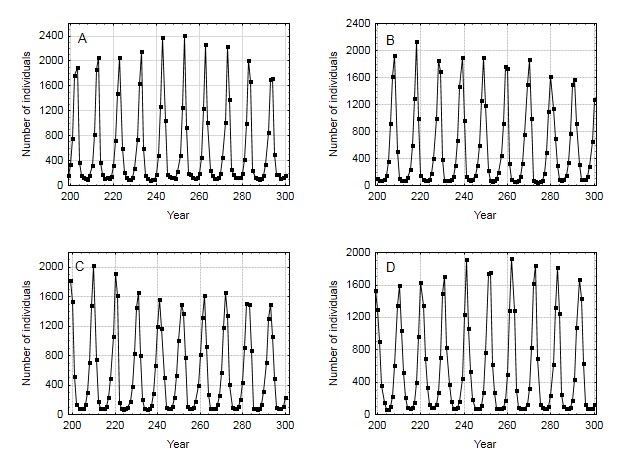}
     \caption{Fully developed outbreak dynamics from year 200 to year 300 for forests with uniform distribution of trees and different sizes. The figure shows the total number of insects in the forest divided by the number of trees in the forest. Forest size: A -- $20\times 20$, B -- $50\times 50$, C -- $75\times 75$, D -- $100\times 100$. The parameters of the model have standard values (Table \ref{table:1}).}
     \label{fig:3}
\end{figure}

\begin{figure} [!h]
    \centering
    \includegraphics[width=\linewidth]{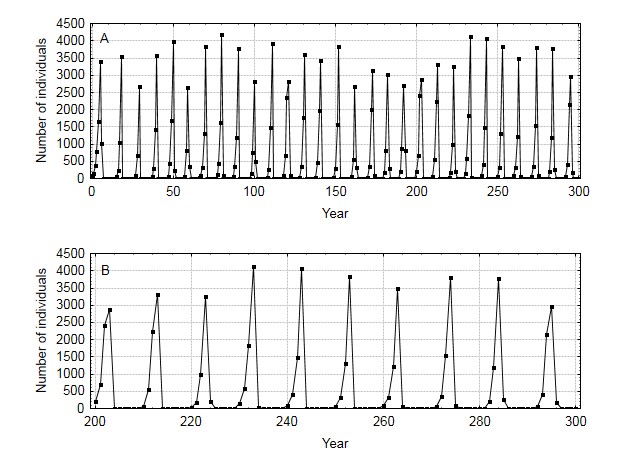}
    \caption{Dynamics of the local population on one tree located in the middle of the forest with uniform distribution of  trees and size $20\times 20$. A -- dynamics over the initial 300 years. B -- dynamics in the range from year 200 to year 300. The parameters of the model have standard values (Table \ref{table:1}).}
\label{fig:4}
\vskip -0.2cm
\end{figure}

\vskip 0.3cm
The dynamics of insect populations in forests with the following sizes: $20\times 20$, $50\times 50$, $75\times 75$ and $100\times 100$ trees was studied. In each of these forests, the trees were uniformly distributed. The results for the initial 300 years are shown in Fig.~\ref{fig:2}. Fig.~\ref{fig:3} presents fully developed population dynamics, i.e.~in the years from 200 to 300. Fig.~\ref{fig:4} illustrates the dynamics of the number of insects inhabiting a tree located in the center of the forest with a size of $20\times 20$ trees, from the colonization of which in the first year of simulation the development of the insect population in the forest began. The local insect population on this tree goes extinct regularly when its resources drop below the threshold value. As the time needed for the tree to regenerate resources passes, it is colonized by insects that migrate from neighboring trees.

\subsection{Uniform distribution of trees, small forests }
\begin{figure} [!h]
        \centering
        \includegraphics[width=\linewidth]{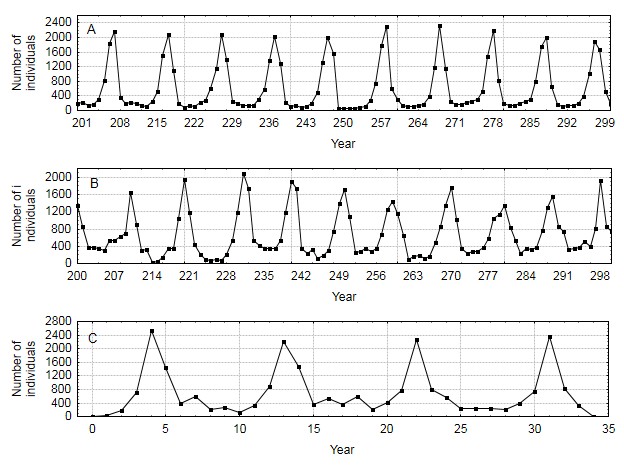}
        \caption{Population dynamics for forests with uniform distribution of trees but small sizes. The figure shows the total number of insects in the forest divided by the number of trees in the forest. A -- $10\times 10$, B -- $5\times 5$, C -- $3\times 3$ (extinction in 34 year). In cases A and B ddynamics in the range from year 200 to year 300 is shown, because populations are persistent in these cases. The parameters of the model have standard values (Table \ref{table:1}).}
        \label{fig:5}
        \vskip -0.2cm
    \end{figure}

\vskip 0.3cm
Simulations were carried out for forests with a uniform distribution of trees, but small sizes. Fig.~\ref{fig:5} shows the dynamics of insect numbers for forests with sizes of $10\times 10$, $5\times 5$ and $3\times 3$ trees. In the case of $10\times 10$ and $5\times 5$ tree forests, the dynamics of the outbreak persisted for 300 years. However, in an extremely small $3\times 3$ forest, the insect population goes extinct after several dozen years. The reason for this is that at some time the resources of all trees drop below the threshold value ${regen}\_{threshold}\, V_0$ and there is no tree on which a~local insect population can persist. In forests that are only slightly larger, such synchronization of tree states is no longer possible, and the insect population persists.

\begin{figure}[!h]
    \centering
    \includegraphics[width=0.75\linewidth]{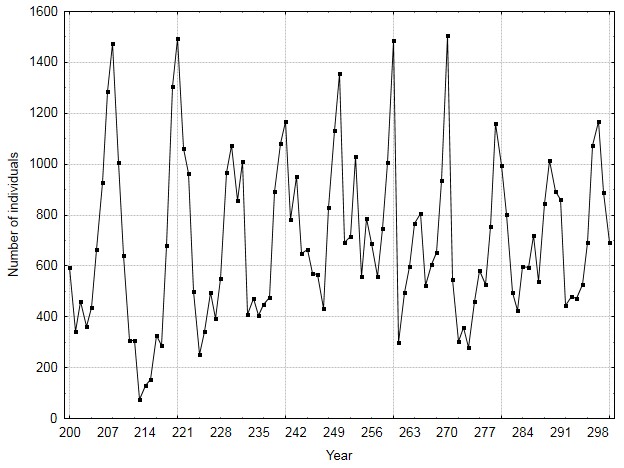}
    \caption{Forest with uniform distribution of trees and a size of $100\times 100$. Population dynamics of insects inhabiting a $5\times 5$ area located in the middle of the forest from year 200 to year 300. The figure shows the total number of insects on these trees divided by 25. The parameters of the model have standard values (Table \ref{table:1}).}
    \label{fig:6}
    \vskip -0.2cm
\end{figure}

Another numerical experiment was performed with a forest with a uniform distribution of trees. The forest with the largest size of $100\times 100$ trees was selected for simulation, but the number of insects on trees from a small area of $5\times 5$ trees located in the middle of the forest was observed. Fig.~\ref{fig:6} shows the changes in the total number of insects on these 25 trees over 100 years from year 200 to 300. This is the period in which the typical outbreak dynamics is fully developed throughout the large forest (see panel D in Fig.~\ref{fig:3}). However, the dynamics of insect numbers in these few trees selected from a large forest hardly resembles those typical of the entire forest.

\subsection{Random distribution of trees, forests with different tree densities}
\begin{figure} [!h]
    \centering
    \includegraphics[width=\linewidth]{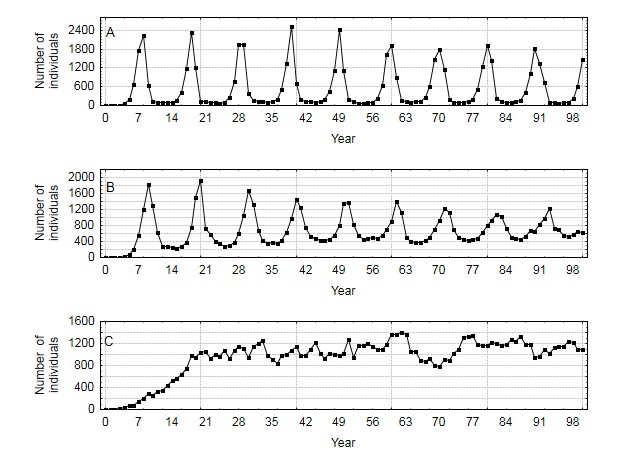}
    \caption{Forest with a random distribution of trees and size $20\times 20$. Population dynamics for different tree densities in the forest. The figure shows the total number of insects  divided by the number of trees in the forest. A -- ${tree}\_{density} = 0.9$, B -- ${tree}\_{density} = 0.5$, C -- ${tree}\_{density} = 0.1$. Parameter ${disp}\_{sd} = 2.5$. The other parameters of the model have standard values (Table \ref{table:1}).}
    \label{fig:7}
    \vskip -0.5cm
\end{figure}

\vskip 0.3cm
Simulations were carried out for a forest of $20\times 20$ trees. The parameter ${tree}\_{density}$ determined the probability that in the initial year a tree with a~resource level equal to $V_{0}$ would be placed in each of the 400 possible tree locations. In this way, forests with random tree distributions, but also with different tree densities, were obtained. For example, if ${tree}\_{density} = 0.1$, this means that the density is 10 times lower than in the initial model with a~uniform distribution of $20\times 20$ trees. Fig.~\ref{fig:7} shows the dynamics of the number of insects in the forest with decreasing values of ${tree}\_{density}$. Simulations were carried out for the first hundred years because in a forest of this size, as can be seen from the results of previous simulations, the type of population dynamics was established very quickly. 

\begin{figure}[!h]
    \centering
    \includegraphics[width=0.75\linewidth]{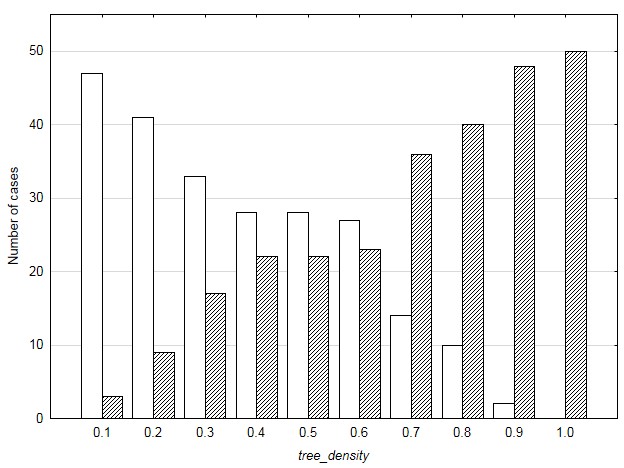}
    \caption{Forest with a random distribution of trees and size $20\times 20$. Distributions of the number of successful insect invasions (solid bars) and the number of failures (empty bars) for 50 repeated simulations for different ${tree}\_{density}$ values. The remaining model parameters have standard values (Table \ref{table:1}).}
    \label{fig:8}
    \vskip -0.2cm
\end{figure}

I also examined the chances of successfully initiating the presence of insects in the forest when, in the initial year, ${N}_{0}$ insects were located in a randomly chosen place in the forest with different tree density. If there was no tree at this location, the insects died in the first time step of the simulation, and the attempt to initiate the presence of insects was unsuccessful. Fig.~\ref{fig:8} shows the distributions of the number of successful insect invasions and the number of failures for 50 repeated simulations for each value ${tree}\_{density}$.

\subsection{Random distribution of trees, different ranges of insect dispersion}
\vskip 0.3cm
A forest with a random distribution of trees and the lowest tree density (${tree}\_{density} = 0.1$) was selected for the simulation. Simulations were carried out for different values of the ${disp}\_{sd}$ parameter: 0.5, 1.0, 1.5, 2.0. Fig.~\ref{fig:9} shows two examples of population dynamics for ${tree}\_{density} = 0.1$ and ${disp}\_{sd} = 0.5$. In a forest with a low density of trees, a population of insects that disperse over short distances has no chance of developing outbreak dynamics. It often goes extinct relatively quickly, and even if it persists over 100 years, it is not an outbreak type of dynamic. In the model presented here, only for values of the parameter ${tree}\_{density}$ greater than or equal to 0.2 the insect population did not go extinct even once out of 100 repetitions, but only for insects with a~large dispersion range (${disp}\_{sd}$ greater than 1.5). 

Fig.~\ref{fig:10} shows the distributions of population extinction times from 100 simulation repetitions for different values of the parameter ${disp}\_{sd}$. In these simulations, it was assumed that a tree originating in the middle of the forest, where the population history begins, always has ${V}={V}_{0}$ at the beginning of the first year. 

\subsection{Uniform distribution of trees, forests with different tree densities }
\vskip 0.3cm
Earlier simulations were carried out for a spatial arrangement in which the square surface was divided into $20\times 20$ patches. Each of them contains a~tree. This corresponded to a $20\times 20$ forest with uniformly distributed trees. It was a tree arrangement scheme in which each tree had 8 neighbors located in the closest distance (i.e. at a distance of one patch). This resulted in maximum tree density. Now, two other distribution schemes have been selected, giving lower tree densities but in each case ensuring a uniform distribution of trees. One fourth of the maximum density (100 trees in the forest) was given by a scheme in which each tree had 8 neighbors located two patches apart. A~density equal to one eighth of the maximum density (49 trees in the forest) resulted in a scheme in which each tree had 8 neighbors, but located at a~distance of three patches. In the latter case, to ensure the same conditions for local populations in trees located near the edge of the forest, the forest space was increased to $21\times 21$ patches.

Fig.~\ref{fig:11} shows examples of insect population dynamics for the two above-mentioned distribution schemes. Simulations were performed for the lowest value of the parameter that characterizes the dispersion of insects (${disp}\_{sd} = 0.5$). Other parameters of the model have standard values. Fig.~\ref{fig:12} shows the dynamics of the local insect population on a single tree located in the middle of the forest, which was inhabited in the initial year of the simulation, for the above two schemes of uniform distribution of trees. 

At a tree density of one fourth of the maximum value, the dynamics of the local insect population on one tree resembles that illustrated in Fig.~\ref{fig:4}. In both cases, the local population grows rapidly and then just as rapidly reaches zero because the level of resources has fallen below the permissible minimum. However, in the case of a density equal to one eighth of the maximum value, the dynamics of the local population is different. After an initial increase, the number of insects begins to fluctuate around a certain level. This is a typical dynamic for this type of single population dynamics model without immigration, which was used in this work (for details, see \cite{uchmanski4}). With such a low tree density and a small dispersion range, individual trees function as isolated populations. In this case, the population also goes extinct when the level of resources falls below the permissible minimum, but this happens very rarely.

\section{Discussion }
\vskip 0.3cm
The dynamics of the population of forest insects of the outbreak type is an obviously emerging feature of the studied system. None of the models' assumptions suggest that the system may behave in this way. This is one of the charms of nature, especially visible when individual-based modeling methods are used to describe the ecological system. Well-known characteristics of individuals, imaginable mechanisms of interactions between individuals, and the effects of these interactions, also largely known from experimental research, when put together do not create a chaotic picture, even though some elements of this system are reproduced in a very large number of copies. Instead, we get a~very orderly behavior of the system that is largely resistant to changes in its operating conditions.

Outbreak dynamics is a common product of the insects' characteristics, but also, perhaps even to a greater extent, of the environment in which the insects live and the way insects behave in this environment. The environment in which insects live consists of a large number of trees. These trees support local insect populations, and insects have the ability to move between these environmental islands. The resources a single tree offers to insects are eaten by the local insect population and regenerated by the tree. If resource consumption is intense (for example, due to the large number of insects feeding on a tree), its amount available locally to insects decreases. If it falls below a certain threshold value, the tree ceases to be a food source for insects and needs some time to regenerate its resources. Insects can behave quite aggressively towards the tree's resources and quickly bring them to such a level that the tree falls out of circulation and requires some time to regenerate (Fig.~\ref{fig:4}).

The outbreak dynamics is an effect of scale. In a sufficiently large forest regardless of the density of trees and their spatial distribution, provided that the range of insect dispersion is large enough, it develops in the form of regular oscillations of number with a very typical shape -- phases of high number and very low number (Fig.~\ref{fig:3}). All factors and events that contain elements of randomness or that act or take place in an uncoordinated way and can lead to more chaotic behavior of the system, with a large number of trees and an even larger number of insects losing their importance, and their effects cancel each other. However, they appear in the case of small forests (Fig.~\ref{fig:5}). The dynamics become more chaotic, we lose the outbreak dynamics, and there is a possibility that the resources of all trees in a small forest will be below the threshold value at the same time and then the insect population in the entire small forest goes extinct. In a large forest, the extinction of insect populations is very unlikely to occur because there is always at least one tree that will provide emigrants. 

\begin{figure}[!h]
    \centering
    \includegraphics[width=\linewidth]{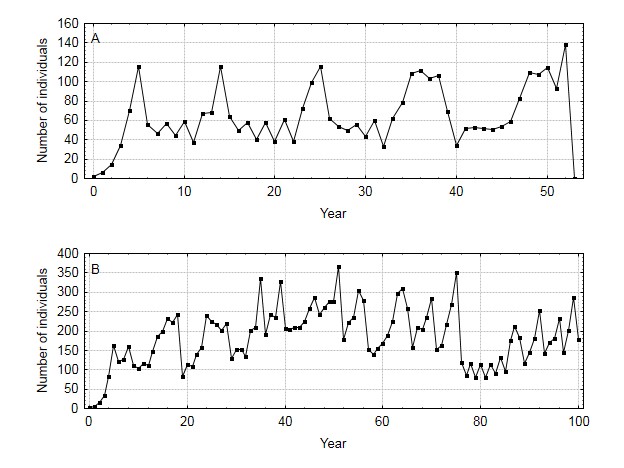}
    \caption{Forest with random distribution of trees and size $20\times 20$. Two examples of population dynamics for ${tree}\_{density} = 0.1$ and ${disp}\_{sd} = 0.5$. The figure shows the total number of insects in the forest divided by the number of trees in the forest. A -- an example of dynamics with an extinction time shorter than 100 time steps. B -- the population does not become extinct in 100 time steps.  The remaining model parameters have standard values (Table \ref{table:1}).}
    \label{fig:9}
\end{figure}
\begin{figure} [!h]
    \centering
    \includegraphics[width=\linewidth]{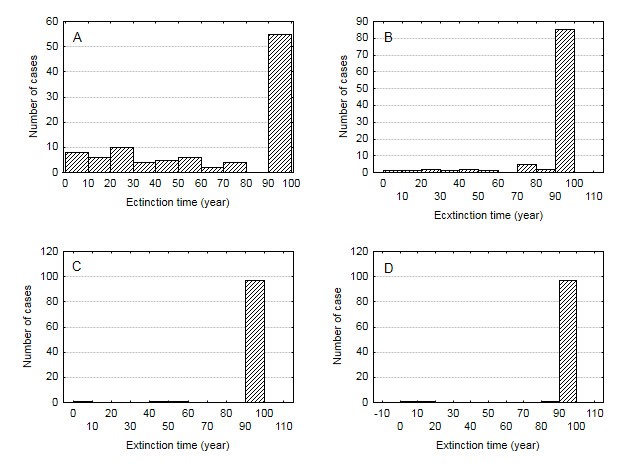}
    \caption{Forest with a random distribution of trees, size $20\times 20$ and tree density determined by the parameter ${tree}\_{density}  = 0.1$. The distributions of the extinction times of populations from 100 simulation repetitions for different values of the parameter ${disp}\_{sd}$ are shown. A -- ${disp}\_{sd} = 0.5$, B -- ${disp}\_{sd} = 1.0$, C -- ${disp}\_{sd} = 1.5$, D -- ${disp}\_{sd} = 2.0$. The remaining model parameters have standard values (Table \ref{table:1}).}
    \label{fig:10}
\end{figure}

This dependence on the scale of the system produces an effect that makes it difficult to detect and study an outbreak of insects in real forests. Observing the dynamics of insect populations in a small subset of neighboring trees from a forest large enough to develop clear outbreak dynamics does not allow us to determine this type of dynamics (Fig.~\ref{fig:6}). We will not even see what is typical of a very small forest (see panel C in Fig.~\ref{fig:5}), because we are now viewing a section of space from a larger system and the observed trees are strongly influenced by migration from neighboring trees.

The full development of outbreak dynamics requires that a sufficient number of trees or all of them are involved in its functioning. A large forest, compared to a small forest, is characterized by a very slow development of outbreak dynamics (Fig.~\ref{fig:2}). The initial phase of population dynamics is not an outbreak dynamics, but an increase in the number of individuals, which ends when all trees in the forest are included in the functioning of the system. At the same time, as the number of infected trees increases, the dynamics of the population begins to resemble an outbreak more and more. If the development of an insect population begins from a single tree, it may take many years to infect all trees and include them in the mechanism that generates outbreaks. During this time, the insects basically remain hidden and the true dynamics of their population in the forest is probably difficult to detect. The time needed for the development of full-scale outbreak dynamics is, of course, correspondingly shorter when the insect invasion in the forest starts from more than one center. In a forest with a uniform distribution of trees, each invasion that starts with one outbreak leads to population expansion. If an insect invasion in a~forest begins with one center, then in a forest with a random distribution of trees it may not develop for the simple reason that insects arriving in the forest first may not end up in the place where the tree grows. The chance of such an event increases as the density of the tree decreases.

\begin{figure} [!ht]
    \centering
    \includegraphics[width=\linewidth]{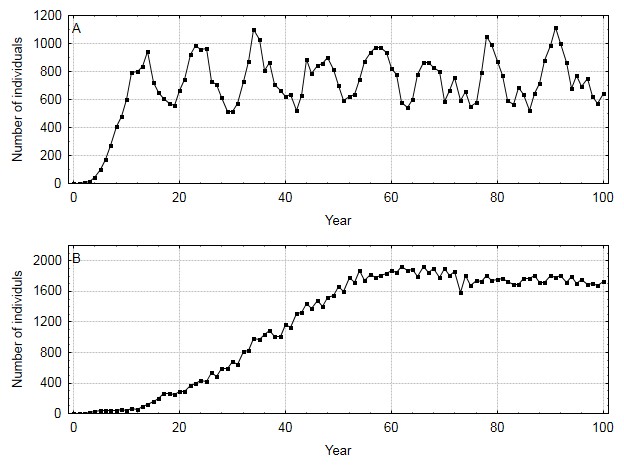}
    \caption{Forest with uniform distribution of trees and size $20\times 20$. Dynamics of insect populations in a forest with different patterns of uniform distribution of trees leading to different tree densities. Parameter ${disp}\_{sd} = 0.5$. The figure shows the total number of insects in the forest divided by the number of trees in the forest. A -- density equal to one quarter of the maximum, B -- density equal to one eighth of the maximum. Other parameters of the model have standard values (Table \ref{table:1}).}
    \label{fig:11}
\end{figure}

It is interesting that the hundred-year sequences illustrating fully developed outbreak dynamics for forests with the same tree density and the same type of tree distribution, but of different sizes, and of course large enough for this type of dynamics to be clearly presented, show the same number of maxima in a given period (Fig.~\ref{fig:3}). This proves that such a feature of the entire system, as the number of all insects in the forest, is shaped by details that are the same for all forests regardless of their size. The period of outbreak is therefore the result of the characteristics of the system's components, such as the life cycle of insects, the nature of interactions between them, and the dynamics of tree resources.

The functioning of this system described in this paper is influenced by three factors: the size of the forest, the density and type of distribution of trees in the forest, and the range of insect dispersion. Outbreak population dynamics will characterize this system regardless of the density of trees and the way they are distributed on the forest surface, as long as the forest is large enough and the range of insect dispersion allows them to move between neighboring trees.

The local dynamics of the number of insects on one tree in a forest, where the dynamics of all insects has the character of an outbreak, is characterized by a~rapid increase in number and then an equally rapid decrease until the extinction of the local population after exceeding the tolerance threshold of the tree (Fig.~\ref{fig:4}). This type of dynamics is primarily the result of changes in the number of individuals living on this tree for generations and their emigration, but the main cause of excessive growth of the local population leading to the tree's tolerance threshold being exceeded is the influx of immigrants from neighboring trees. There are usually more immigrants from neighboring trees in the local population than there are emigrants from that population. This happens because each tree has several neighbors. Between these phases of growth and decline, the number of locals is zero. These local events are synchronized to some extent by insects migrating to neighboring trees, which is why the dynamics of the entire insect population in the forest goes through phases of similar oscillations. What is noteworthy is approximately the same period of oscillation of numbers in the entire forest and on one tree. The difference is only that at the maximum the number of insects in the entire forest (calculated for an average tree) is smaller than in the same phase in one tree, because at each year there are trees in the forest not inhabited by insects, and at the minimum the number does not reach zero, because there are always trees inhabited by insects.

\begin{figure} [!h]
    \centering
    \includegraphics[width=\linewidth]{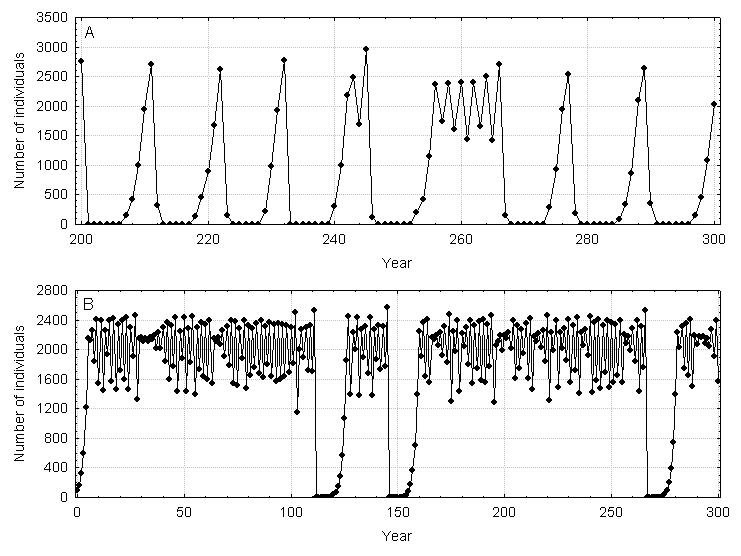}
    \caption{Dynamics of the local population on one tree located in the middle of the forest. A forest with uniform distribution of  trees and size $20\times 20$. Parameter ${disp}\_{sd} = 0.5$. A -- density equal to one quarter of the maximum. Dynamics in the range from year 200 to year 300. B -- density equal to one eighth of the maximum. Dynamics over the initial 300 years. Other parameters of the model have standard values (Table \ref{table:1}).}  
    \label{fig:12}
\end{figure}

However, when the range of insect dispersion is small and/or the density of trees is low, the system stops functioning in the way described above. It becomes a set of more or less isolated local populations that influence each other in a very limited way. The dynamics of the local population, especially for the forest with a uniform distribution of trees, begins to resemble the type of dynamics that is characteristic of the single population model with global competition used in this work to construct the model of forest insect outbreak \cite{uchmanski4}. It turns out that the type of dynamics of a single population is significantly dependent on the dynamics of the resources. In the case of the model used here, we have resources with limited growth of the logistic type. In such a case, the isolated, local population will initially be characterized by an exponential increase in size, and later it will settle at a certain level around which it will fluctuate. This type of local population dynamics in a single tree is illustrated in panel B in Fig.~\ref{fig:12}.

When the dispersion range is large enough, individuals are exchanged between neighboring trees. Then the above type of local population dynamics on a single tree disappears (see panel A in Fig.~\ref{fig:12}). Immigrants significantly increase the local population. This leads to intensive exploitation of tree resources, and the population quickly goes extinct after the initial growth period. This type of local population dynamics is characteristic of the outbreak dynamics of insect populations in the entire forest (Fig.~\ref{fig:3}).

The type of tree distribution in the forest is important for the dynamics of insect population. However, it only becomes visible when the density of trees becomes low and/or the range of insect dispersion is small. In general, it can be said that the insect population in the forest with randomly distributed of trees becomes more susceptible to extinction (Fig.~\ref{fig:9}). This will happen because in a forest with a random distribution of trees, one can expect in some parts of the forest large distances between neighboring trees where the exchange of insects between trees is impossible and in other parts local clusters of trees can be expected, where the exchange of insects between neighboring trees, with its negative effects on the local population, will be possible. When there is a higher density of randomly distributed trees and sufficiently large range of insect dispersion, the dynamics has the outbreak character. As the density decreases, the amplitude of oscillations in the number of insects decreases in subsequent time steps, and for small densities it transforms into dynamics of a permanent outbreak (Fig.~\ref{fig:7}). The important thing is that this is the result of the development of insect populations throughout the forest -- all trees are involved in this process. For very low tree densities in the forest, fluctuations in the number of insects in the forest become increasingly chaotic. This effect is further enhanced by reducing the range of insect dispersion. In this case, only a small number of trees support local populations of insects. Repeated simulations show that, in addition to situations where the population persists throughout the simulation period, there are now cases of earlier extinctions of insect populations throughout the forest (Figs.~\ref{fig:9} and \ref{fig:10}). In contrast, a~forest with low density of trees but with an uniform distribution of trees will function as a sum of more or less isolated local populations without the negative impact of immigrants on the dynamics of these populations (Fig.~\ref{fig:11}).
\vskip -0.5ex
Individual-based models allow for linking the features of the described ecological system with the features of individuals and the environment in which they live. This provides a rational basis for using information about the biology of species that comes, for example, from laboratory studies. On the other hand, it allows us to predict changes in the behavior of an ecological system if we modify the characteristics of individuals. In the model presented above, it was assumed that it described a parthenogenetic insect species. The simplest way to introduce sexual reproduction would be to assume that one half of the adults that emerge after the larval growth period are males and the other half are females. Copulation takes place on the tree where pupation occurred. Only females migrate to neighboring trees. Since we are dealing with larvae that differ in weight, it would be easiest to assume that there is no sexual dimorphism and classify individuals into each sex with the same probability of 0.5. This could be compared to the effects of reducing offspring production in a model with a parthenogenetic species, which, as we know from previous analyzes \cite{uchmanski3}, does not exclude the occurrence of mass outbreaks, but only increases the chance of permanent outbreak solutions. These effects could be eliminated if we assumed increased female dispersion. Similarly, one may wonder about the effects of temperature changes, which will affect primarily the course of physiological processes. An increase in temperature will certainly increase the respiration rate, which may result in a reduction in the weight of individuals. In the adopted scheme of system operation, this will not have a major impact on its dynamics. However, the increase in temperature will undoubtedly affect the rate of primary production, which may increase the rate of tree regeneration, which in turn will shorten the period of mass outbreaks \cite{uchmanski3}.

\acknowledgments{The author thanks anonymous reviewers for improving the quality of~the paper.}
\vskip 0.5cm

\centerline{\large\bf References}


\begin{thebibliography}{99}
\bibitem{balten}W.~Baltensweiler, G.~Benz, P.~Bovey and  V.~Delucchi, Dynamics of larch bud moth populations, Annual Review of Entomology \textbf{22}, 77--100 (1977), \doi{10.1146/annurev.en.22.010177.000455}.
\bibitem{barbosa}P.~Barbosa and J.C.~Schultz (Eds), \emph{Insect Outbreaks}, Academic Press (1987).
\bibitem{berryman}A.A.~Berryman, \emph{Forest insects. Principles and Practice of Population Management}, Plenum Press (1986), \doi{10.1007/978-1-4684-5080-4}.
\bibitem{berryman2}A.A.~Berryman (ed.), \emph{Dynamics of Forest Insect Population. Patterns, Causes, Implications}, Plenum Press (1988), \doi{10.1007/978-1-4899-0789-9}.
\bibitem{duncan}A.~Duncan  and R.Z.~Klekowski,  Parameters of energy budgets. In: W.~Grodziński, R.Z.~Klekowski  and A.~Duncan (Eds), \emph{Methods of Ecological Bioenergetics}, Blackwell Scientific Publications, 97--147 (1975)
\bibitem{hassel}D.~Hassell, D.~Allwright  and A.~Fowler,  A mathematical analysis of Jone’s site model of spruce budworm infestation, Journal of Mathematical Biology \textbf{38}, 377--421 (1999) \doi{10.1007/s002850050154}.
\bibitem{isaev}A.S.~Isaev, R.G.~Khlebopros, L.V.~Nedorezov, Yu.P.~Kondakov, V.V.~Kiselev, V.G.~Soukhovolsky, \emph{Populacjonnaja Dinamika Lesnyh Nasekomyh}, Nauka (2001) (in Russian).
\bibitem{isaev2}A.S.~Isaev, R.G.~Khlebopros, V.V.~Kiselev,  Yu.P.~Kondakov, L.V.~Nedorezov,  V.G.~Soukhovolsky, Forest insect population dynamics, Euroasian Entomological Journal \textbf{8}, suppl. 2, 1--115 (2009), \doi{10.1002/9781119407508}.
\bibitem{ivlev}V.S.~Ivlev, \emph{Experimental Ecology of the Feeding of Fishes}, Yale University Press (1961).
\bibitem{kolk}A.~Kolk, L.~Sukovata and  M.~Dobrowolski, \emph{Warunki Gradacji Owadów w Pierwotnych Ogniskach Gradacyjnych na Przykładzie Puszczy Noteckiej i Borów Tucholskich}, Instytut Badawczy Leśnictwa (2005) (in Polish).
\bibitem{ludwig}D.~Ludwig, D.D.~Jones and C.S.~Holling, Qualitative analysis of insect outbreak system, the spruce budworm and forest, Journal of Animal Ecology \textbf{74}, 315--332 (1978), \doi{10.2307/3939}.
\bibitem{ludwig2}D.~Ludwig., D.G.~Aronson and  H.F.~Weinberger, Spatial pattern of the spruce budworm, Journal of Mathematical Biology \textbf{8}, 3, 217--258 (1979), \doi{10.1007/BF00276310}.
\bibitem{lomnicki}A.~Łomnicki, \emph{Population Ecology of Individuals}, Princeton University Press (1988), \doi{10.2307/j.ctvx5wbhx}.
\bibitem{morris}R.F.~Morris  (Ed.), The dynamics of epidemic spruce budworm populations, Memoires of the Entomological Society of Canada \textbf{21} (1963), \doi{10.4039/entm9531fv}.
\bibitem{nedorezov}L.V.~Nedorezov,  \emph{Modelirovanije Vspyshek Massovyh Rozmnozenij Nasekomyh},  Nauka (1986) (in Russian).
\bibitem{shu}H.~Shu, W.~Xu, X.~Wang  and J.~Wu, Spatiotemporal patterns of a structured spruce budworm diffusive model, Journal of Differential Equations \textbf{336}, 427--455 (2022), \doi{10.1016/j.jde.2022.07.014}.
\bibitem{uchmanski}J.~Uchmański, Differentiation and frequency distributions of body weights in plants and animals, Philosophical Transactions of Royal Society London Ser. B \textbf{310}, 1--75 (1985).
\bibitem{uchmanski2}J.~Uchmański, Resource partitioning among unequal competitors, Ekologia Polska \textbf{35}, 71--88 (1987). 
\bibitem{uchmanski3}J.~Uchmański, Cyclic outbreaks of forest insects: a two dimensional individual-based model, Theoretical Population Biology \textbf{128}, 1--18 (2019), \doi{10.1016/j.tpb.2019.04.006}.
\bibitem{uchmanski4}J.~Uchmański, Catastrophic dynamics: how to avoid population extinction? Ecological Modelling \textbf{483} (2023), \doi{10.1016/j.ecolmodel.2023.110438}.
\bibitem{uchmanski5}J.~Uchmański and J.~Dgebuadze,  Factors effecting skewness of weight distributions in even-aged populations: a numerical model, Polish Ecological Studies \textbf{16}, 297--311 (1990).


\end{thebibliography}
\bibliographystyle{abbrv}

\Koniec
\end{document}